\documentclass[12pt]{article}
\usepackage{url, aas_macros, natbib, graphicx, wrapfig,amssymb}
\usepackage[top=1.0in, left=1.0in, right=1.0in, bottom=1.5in]{geometry}
\usepackage[nottoc,numbib]{tocbibind}

\newcommand{\captionfonts}{\small}
\makeatletter  
\long\def\@makecaption#1#2{%
  \vskip\abovecaptionskip
  \sbox\@tempboxa{{\captionfonts #1: #2}}%
  \ifdim \wd\@tempboxa >\hsize
    {\captionfonts #1: #2\par}
  \else
    \hbox to\hsize{\hfil\box\@tempboxa\hfil}%
  \fi
  \vskip\belowcaptionskip}
\makeatother   

\begin{document}

\title{Targeting Young Stars with Kepler: Planet Formation, Migration Mechanisms and the Early History of Planetary Systems}
\author{James P. Lloyd,  Jonathan I. Lunine (Cornell University),\\
Eric Mamajek (University of Rochester),\\
David S. Spiegel (Institute for Advanced Study),\\
Kevin R. Covey, Evgenya L. Shkolnik (Lowell Observatory),\\
 Lucianne Walkowicz (Princeton University),\\ 
 Miguel Chavez, Emanuele Bertone, Jose Manuel Olmedo Aguilar\\
  (Instituto Nacional de Astrof\'{i}sica, \'{O}ptica y Electr\'{o}nica)}
\date{}
\maketitle

\setcounter{page}{1}

\parskip = 10pt
\parindent = 0pt

\begin{abstract}

This white paper discusses a repurposed mission for the Kepler spacecraft that focusses on solving outstanding problems in planet formation and evolution by targeting the study of the hot Jupiter population of young stars.  This mission can solve the question of the mode of migration of hot Jupiters, address the problem of whether Jupiters form by hot-start (gravitational instability) or cold-start (core accretion) mechanisms, and  provide a wealth of data on the early stages of planetary system evolution during the active phases of stars which  impact planetary habitability.  In one year of observations of three weeks dwell time per field, Kepler would increase by more than an order of magnitude the number of known hot Jupiters, which can be followed up with fast cadence observations to  to search for transit timing variations and to perform asteroseismological characterization of the host stars.  This mission scenario continues to operate Kepler in the photometric monitoring mode for which it was designed, and is generally flexible with regards to field selection enabling prioritization of fuel usage and attitude control constraints.

\end{abstract}

\newpage

\section{Young Stars: Probes of Formation and Hot Jupiter Migration}

A measurement of the rate of hot Jupiters around stars younger than 1 Gyr will discriminate between candidate migration scenarios that form hot Jupiters. 
There are two classes of formation mechanism to explain the existence of hot Jupiter planets: migration by various mechanisms in the protoplanetary disk (e.g. \citet{Alibert:2004il,Alibert:2005jl,Bodenheimer:2000fu,Bryden:1999bv,Fabrycky:2007pt,Ida:2004xw,Johnson:2006qq,Kley:2000kb,Laughlin:2004zt,Levrard:2009hq,Lin:1996fe,Mandell:2007cq,Masset:2003tw,Menou:2004ee,Moorhead:2005gb,Murray:1998lh,Nelson:2000qf,Papaloizou:2000ye,Raymond:2006xq,Thies:2011rq,Trilling:1998ec,Trilling:2002qo,Udry:2003jw,Ward:1997ff,Wu:2003os})
and a variety of  mechanisms that dynamically produce  highly eccentric orbits and evolve the orbit through tides (e.g. \citet{Albrecht:2012rr,Fabrycky:2009ul,Jackson:2008ve,Lai:2011bh,Nagasawa:2008gf,Naoz:2011lq,Naoz:2012dq,Rasio:1996mz,Triaud:2011pd,Winn:2010qf}).  Approaches to discriminate statistically between these scenarios have been proposed \citep{Armitage:2007ss,Schlaufman:2010yo,Watson:2011qc}, but the statistical tests are not strong and confounded by completeness and systematic effects when comparing transit surveys and radial velocity surveys.  However, the two scenarios deliver the hot Jupiters at differing times relative to the formation of the star: promptly in the case of protoplanetary disk migration, and delayed by the tidal evolution timescales of $>$0.5 Gyr.  Therefore the abundance of hot Jupiters around a sample of stars younger than 1 Gyr relative to the abundance of hot Jupiters around a sample of old stars (e.g. the radial velocity survey samples that are selected for low $v \sin i$ and jitter) will provide a powerful test of the hot Jupiter formation mechanism.

The Kepler mission has delivered a flood of exoplanet candidates and stellar astrophysics discoveries garnering wide attention for followup, both from within the Kepler-project organized follow-up program and  in the wider community.   The target selection for the Kepler 
project is primarily based on photometric colors in the SDSS $g$, $r$, $i$, and $z$ bands \citep{Brown:2011ij}, and does not explicitly select based on age indicators or temporal properties of the stars.     Young stars are variable and so are preferentially removed from Kepler monitoring if too variable to detect an Earth sized planet, but Jupiter-sized planets can be detected for many stars with variability well beyond this threshold.   Planets transiting young stars are particularly compelling, as diagnostics on the formation mechanisms of planets are lost as planets age and cool.  Identification of a even a single young ($<$ 100 Myr) planet for which the radius and age could be precisely measured would provide direct insight into the formation mechanism: gravitational instability forms optically thick objects that cool slowly, whereas core accretion is expected to dissipate a larger fraction of the binding energy in an accretion shock \citep{Marley:2007dk}.

\begin{figure*}[h]
\includegraphics[width=6in,angle=0]{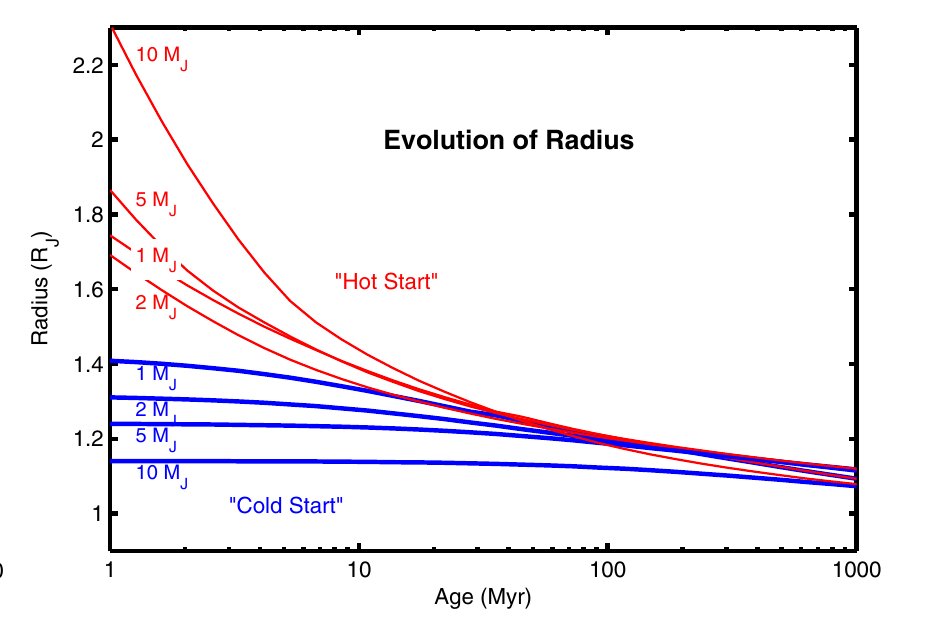}
{\caption { \small Radius-age relations for "hot-start" vs "cold start" formation models of extrasolar planets \citep{Spiegel:2012kx}.}
\label{fig:Spiegel}
}
\end{figure*}

Based on analysis of UV excess determined from our GALEX survey of the Kepler field, we estimate it is unlikely that the existing Kepler dataset will be sufficient to answer these important questions in exoplanet science.   In particular the existing Kepler archival data does not contain enough of the youngest stars to answer definitively the question of the formation mechanism of Jupiters.
 
Although  variability and spot-induced radial velocity shifts will present challenges for both the detection of transit signals and followup confirmation, Jovian planets, which present a large transit ($\sim$1\%) and radial velocity ($>$ 50 m/s) amplitudes are worth pursuing.  Even in the presence of large-amplitude stochastic variability, the uniqueness and repetition of a 1\% transit signal is detectable given sufficient photometric precision.

 \section{Observing Strategy}
 
Generally, there are now two possible observing strategies with Kepler that are consistent with the basic planned mode of operation of the observatory (i.e., to stare at a field on the sky for an extended period of time) that do not run into the limitations of the telemetry budget or substantial modifications to the spacecraft or analysis software.

\begin{enumerate}
\item Continue looking at the same field, just with degraded precision, and learn about the frequency of Jupiters at $\gtrsim$ 1AU.\\

\item Look at many different fields for a much shorter time each.

\end{enumerate}

While option (1) holds some appeal, radial velocity searches have already probed the statistics of Jupiter-mass objects at $\sim$1 AU separations, and so we believe that, since Kepler can no longer perform the 10-100 $\mu$mag photometry needed to find Earths around Sun-like stars, the science yield may be significantly enhanced by pursuing some version of option (2).

By looking at many fields on the sky for $\sim$3 weeks each, Kepler can dramatically increase the number of known hot Jupiters.  Currently, the ground-based transit observatories (primarily, the HAT surveys and SuperWASP) have surveyed of order 1-2 million stars, roughly an order of magnitude more stars than Kepler (see argument and references in \citet{Metzger:2012fk}).  HAT and SuperWASP have announced roughly 110-125 planets (http://www.exoplanets.org or http://www.exoplanet.eu), whereas in the first 13 quarters, Kepler found approximately 90 candidates with orbital periods less than 7 days and radii greater than half of Jupiter's, most of which would have been discovered in just 3 weeks of operation with 1-10 mmag photometry.

By increasing the number of Kepler-sized fields that have been searched via Kepler from 1 to nearly 20 in a year of operation with our recommended observing strategy, Kepler can scan 5\% of the sky and, in so doing, increase by more than an order of magnitude the number of known hot Jupiters.   In each field, we recommend observing 100,000 stars at 20-minute cadence, slightly increasing the observing rate and decreasing the number of targets because of the reduced photometric precision.  Since Kepler's selection function is much better understood than that of HAT and SuperWASP, it will be easier to assess the statistics of the underlying population.   In addition, many of the stars will be nearby and amenable to radial-velocity followup and infrared atmospheric characterization.  Finally, a concomitant benefit is that we might learn whether the statistics of hot Jupiters varies significantly with sight line through the Galaxy.  In this manner, Kepler can approximate the search strategy of the Transiting Exoplanet Survey Satellite (TESS) before its launch.

In a second year of operation, Kepler could either continue the same search strategy on another 17 fields or could revisit the fields of the first year of 2-wheel mode and could, in each field, use fast cadence (1-minute) mode to observe all detected candidates.  As long as there are fewer than 5000 candidates per field (which would surely be the case, given just 3 weeks per Þeld), this would not pose a telemetry bottleneck.  Our recommendation is to revisit the fields of Year-1 of 2-wheel mode and observe the candidate-hosting stars at faster cadence, but this decision could be modified if there is a compelling reason to do so.  Fast-cadence mode would allow searches for transit-timing variations, indicating the presence of other objects in the system.  Furthermore, it would allow asteroseismic characterization of the structure and the evolutionary state and age of (some of) the stars.

\subsection{Possible Field Selections}

Any sight line in the Galaxy contains a mix of old and young stars, and so there are many suitable fields.  Fields including nearby stellar associations with especially young or well determined ages are advantageous although not absolutely required (see discussion below).  For the purposes of this white paper, we have not addressed explicitly the question of selection of fields, concluding that there sufficient possibilities for field selection that the optimization of the attitude control and fuel usage requirements can take priority to enable the longest possible ongoing mission.  This science can be accomplished with continued monitoring of the Kepler field, but with a focus on the youngest stars, although it is better served by increasing the size of the sample by monitoring a larger number of fields for a few weeks at a time.

\vspace{-2mm}
\subsubsection{The Number of Young Stars Accessible in Field Pointings}
\vspace{-2mm}

Assuming a homogenous sample of stars, we may naively estimate that for a constant star formation rate, the youngest 2\% of stars are younger than 2\% of the age of the Galaxy.  Therefore it should be possible to select a sample of order 3000 stars aged less than 250 Myr with comparable magnitude limits to a nominal Kepler long cadence planet search sample of 150,000 stars in a single field with $r \lesssim 16$ mag.  A sample of 3000 stars is  sufficient for a likely detection of at least one transiting Jupiter, if the rate of hot Jupiters is comparable in this sample to the $\sim$ 1\% rate for sun-like stars \citep{Wright:2012vn}.

\begin{figure}
\begin{center}
\includegraphics[width=0.9\textwidth]{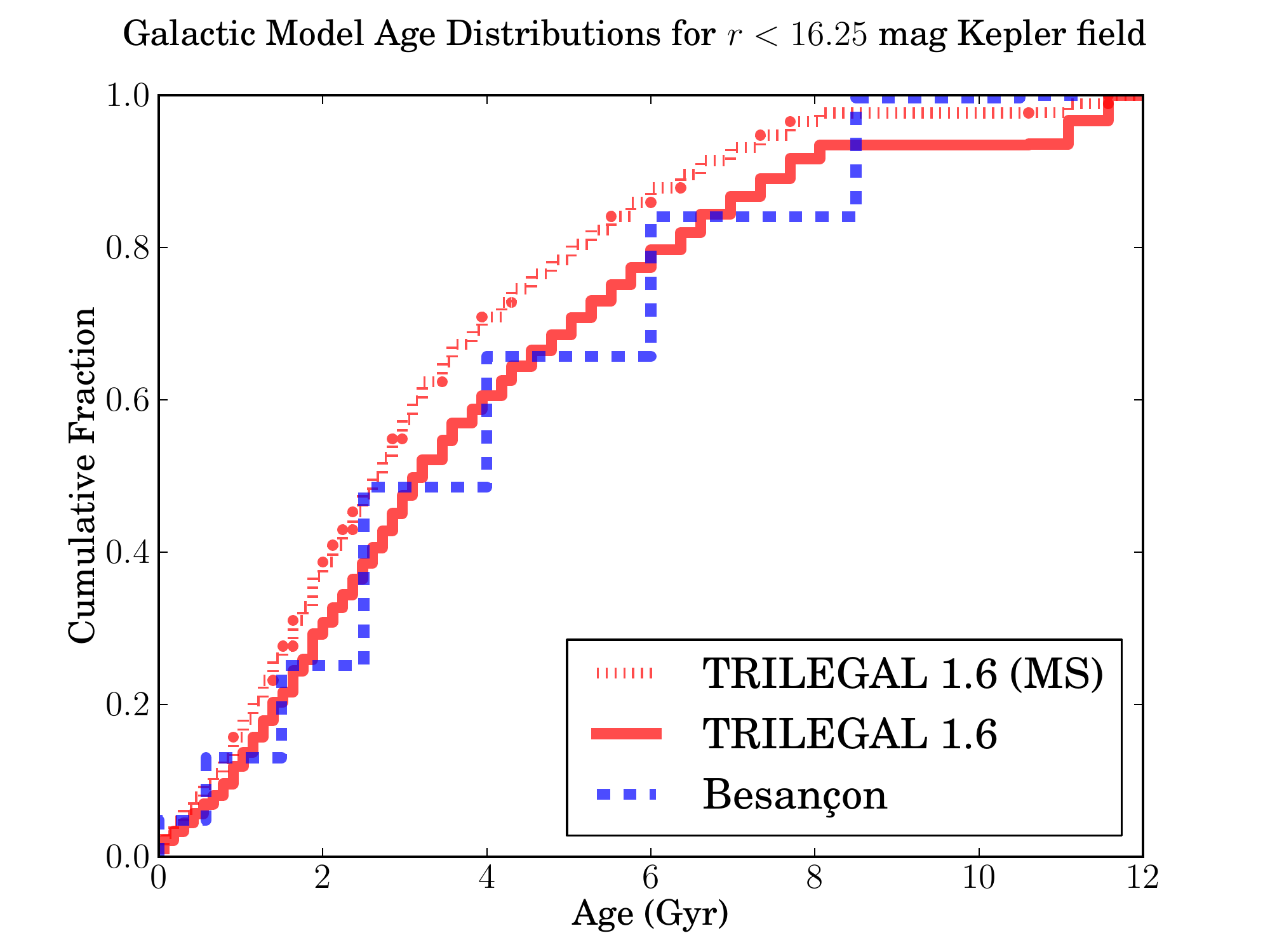}
\caption{\small Age distribution for a $r<16.25$ magnitude limited sample for 10 deg$^2$ in the direction of the Kepler field center.  The TRILEGAL (in red) and Besan\c{c}on (in blue) models are in agreement.  The TRILEGAL model output is has also been separated into the main sequence component (75\% of the total), which is a slightly younger population on average than the strictly magnitude limited sample.}
\label{fig:agedist}
\end{center}
\end{figure}

Galactic dynamics results in a larger scale-height for older stars, altering the space density of stars of a given age as a function of height above the planet of the Galaxy.  Both the luminosity and lifetime of stars are a function of stellar mass so Malmquist bias will not necessarily affect populations of disparate ages comparably.   As a concrete example, which is broadly applicable to comparable lines of sight with moderate galactic latitude, we have assessed the distribution of ages of stars in the Kepler field using Galactic population synthesis models.  Since the young stars will be concentrated primarily in the mid plane of the Galaxy, whereas the Kepler line of sight lies well above the mid-plane at large distances, at fainter magnitudes there will be fewer young stars.  To refine the estimates of accessible young stars beyond the naive assumption above we have calculated the age distribution of stars with a  magnitude limit suitable for Kepler photometry, $r<16.25$, using TRILEGAL 1.6 \citep{Girardi:2012uq} and Besan\c{c}on \citep{Robin:2003fk}.  In Figure~\ref{fig:agedist}, we show Galaxy models for the Kepler field.  Other fields with comparable galactic latitude will show comparable distributions.  The two models have no significant disagreement and estimate that the youngest 2\% of  $r<16.5$ mag stars in the Kepler field are 100 Myr old or less, 7\% are younger than 0.5 Gyr and 15\% are younger than 1 Gyr.

\subsubsection{Target Selection}

 Improved  statistics can be obtained by expanding the sample size and tailoring the selection of targets to young stars using selection criteria such as UV excess.  
Stars arrive on the Main Sequence rotating rapidly, but then spin down with age as they lose angular momentum via braking by magnetized stellar winds.   Activity indicators, particularly UV excess, are widely employed to identify young stars.  Young stars are typically photometrically variable, and therefore make poor targets for transit searches, particularly for small planets.  However, searching for planets specifically orbiting young stars is compelling.   Identification of transiting young planets would provide significant constraints into the dichotomy of hot-start vs cold-start planet models \citep{Marley:2007dk} and so there have been substantial efforts 
devoted to young-star planet searches with ground-based transit surveys  \citep{Aigrain:2007kl,Neuhauser:2011qa,van-Eyken:2011rw}, with mixed and exciting but ambiguous success \citep{van-Eyken:2012mi,Barnes:2013uq}.  These ground-based surveys face serious challenges in addition to the intrinsic variability of the stars: photometric errors comparable to the signal being searched for; incomplete and irregular temporal sampling; and substantial systematic photometric errors.

In practice, the age determination of any individual star cannot be expected with great accuracy, especially based on a single criterion such as UV excess.  Ideally, the complete package of UV excess, photometric variability and rotation period (which will be available from the Kepler time series), Ca H\&K indices and  $v\sin i$ (which would be available from confirmation follow-up radial velocity spectroscopy, would provide the most accurate age estimates.  However, UV excess is the only age indicator that is available for a large number of stars.  The robustness of UV excess selection criteria is sufficient that a sample can be selected that is dominated by stars with ages less than 1 Gyr or younger.  

\begin{figure}
\begin{center}
\includegraphics[width=\textwidth]{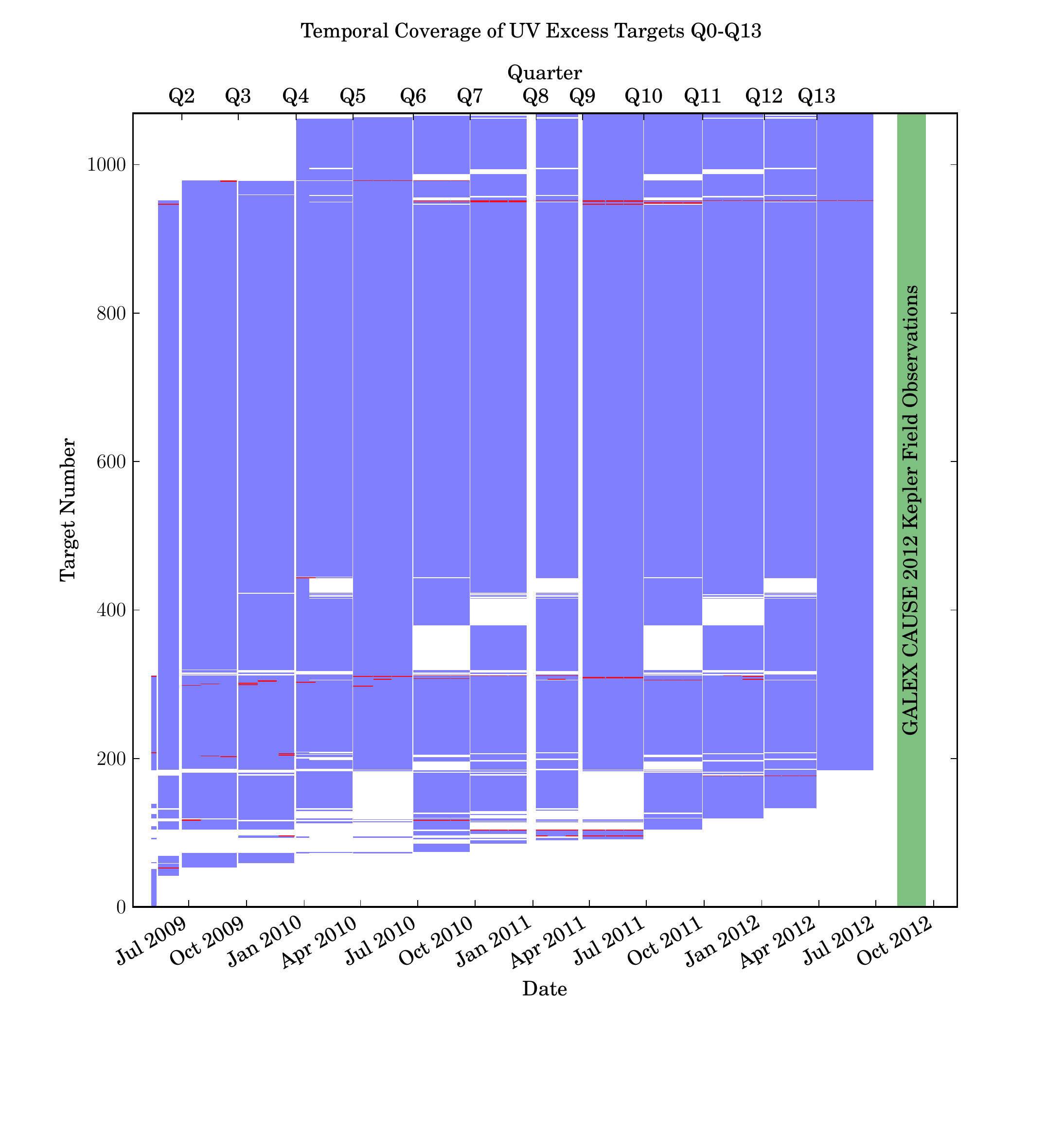}
\caption{Archived  Kepler observations of the 1069 sources in our young star sample (determined by cross-matching with GALEX observations of the Kepler field).   Long cadence photometry are shown in blue and short cadence photometry in red.  50 of these targets have been dropped since Q1, and at least 900 have been observed for 8 quarters.  The GALEX 2012 Kepler field campaign (shown in green) was contemporaneous with Kepler Q14.}
\label{fig:prevbar}
\end{center}
\end{figure}

If the UV excess were perfectly monotonically correlated with age, it would be possible to select a sizable sample of very young stars, since the youngest 2\% of stars accessible to Kepler are therefore  younger than 100 Myr and should be expected to show young Jupiters with a measurable inflation from their formation, if such objects exist.   It is unlikely, however, that the UV excess selection is perfect, but we should be confident that this UV excess selected sample can be reliably dated in bulk to be no older than 0.5-1 Gyr, which is young enough to show clearly whether or not hot Jupiters form promptly, or in a delayed scenario.   Even though there  is  clearly a substantial rate of false positives, even in the most extreme estimate of the false positive rate  (e.g. \citet{Santerne:2012kx}), it remains the case that {\em most} of the unconfirmed KOI are not false positives, and the statistical corrections for false positive rates are modest, even though individual objects remain uncertain without followup.   Kepler data can provide accurate rotation periods, and age estimators can be cross-calibrated with the asteroseismic and rotation data for thousands of stars already observed with Kepler.  It is reasonable to expect that the stars in this survey can be age dated in bulk to the $\sim 1$ Gyr precision necessary to accomplish the statistical determination of the hot Jupiter migration scenario.   Although obtaining  ages  to a precision better than a few hundred Myr for these stars will be difficult in the immediate future, GAIA will obtain accurate space motions, which will produce reliable kinematic ages, especially for the youngest populations.  Finally,  reliable Kepler photometry will contribute to improving our understanding of the age-activity-rotation relation and therefore our ability to estimate the ages of stars in general.

\section{Conclusion}

The use of Kepler to continue to search for planets with reduced precision necessarily implies targeting larger planets than the Kepler prime mission.   These observations would further support the Kepler prime mission by bringing a deeper understanding of planetary systems in general.  In particular the question of whether the migration of Jupiters occurs early or late impacts the stability and habitability of smaller planets.  By targeting young stars in multiple fields, a large sample of hot Jupiters can be built up and used to discriminate between planetary formation and migration scenarios.   This mission scenario is flexible with respect to the choice of fields and therefor allows the optimization of the attitude control to preserve critical fuel and therefore mission lifetime.

\end{document}